\documentclass[twocolumn,floatfix,pra,aps,showpacs]{revtex4}
\usepackage{epsfig,graphicx}
\usepackage{flafter}
\usepackage{amsmath}
\usepackage{color}

\begin{document}
\title{
Phase-slippage and self-trapping in a self-induced bosonic Josephson junction}
\author{M. Abad}
\affiliation{Departament d'Estructura i Constituents de la Mat\`{e}ria,\\
Facultat de F\'{\i}sica, Universitat de Barcelona, E--08028 Barcelona, Spain}
\author{M. Guilleumas}
\affiliation{Departament d'Estructura i Constituents de la Mat\`{e}ria,\\
Facultat de F\'{\i}sica, Universitat de Barcelona, E--08028 Barcelona, Spain}
\author{R. Mayol}
\affiliation{Departament d'Estructura i Constituents de la Mat\`{e}ria,\\
Facultat de F\'{\i}sica, Universitat de Barcelona, E--08028 Barcelona, Spain}
\author{M. Pi}
\affiliation{Departament d'Estructura i Constituents de la Mat\`{e}ria,\\
Facultat de F\'{\i}sica, Universitat de Barcelona, E--08028 Barcelona, Spain}
\author{ D. M. Jezek   }
\affiliation{IFIBA-CONICET and Departamento de F\'{\i}sica, FCEN-UBA\\
Pabell\'{o}n 1, Ciudad Universitaria, 1428 Buenos Aires, Argentina}

\date{\today}

\begin{abstract}

A dipolar condensate confined in a toroidal trap constitutes a self-induced Josepshon junction when the dipoles are oriented perpendicularly
to the trap symmetry axis and the $s$-wave scattering length is small enough. The ring-shaped double-well potential coming from the
anisotropic character of the mean-field dipolar interaction is robust enough to sustain self-trapping dynamics, which takes place when
the initial population imbalance between the two wells is large. We show that in this system the self-trapping regime is directly related to
a vortex-induced phase-slip dynamics.
A vortex and antivortex
are spontaneously  nucleated in the low density regions, before a minimum of the population imbalance is reached, and then cross the toroidal section in opposite directions through the junctions.
This vortex dynamics yields a
phase slip between the two weakly linked condensates causing
an inversion of the particle flux.

\end{abstract}


\maketitle

Phase-slippage is a dynamical process that generally takes place in ac Josephson junctions. In this regime the phase difference increases linearly in time and proportionally to the external voltage applied or chemical potential difference, respectively in superdonductor or superfluid junctions. 
This linear growth can be interpreted as a periodic change of the phase difference by an amount of $2\pi$, or as the system periodically undergoing a phase-slip.
The current understanding is that phase slips are related to the dynamical creation of vortices, which cross the flow path and leave a $2\pi$ phase behind \cite{Anderson1966}.
This process has been widely addressed in superfluid helium (see, for instance, \cite{Donnelly}) and is receiving increasing attention in Bose-Einstein condensates (BECs) \cite{Piazza2009,Piazza2010,Ramanathan2010}.

In the field of superfluid He it has been shown that after the phase-slip takes place the superfluid velocity is reduced by a quantized amount, and under some conditions the flow of atoms can even be reversed.
The conditions for reaching the phase-slip regime depend strongly on the geometry of the constriction, that is the weak link that creates the Josephson junction.
When the width of the junction is larger than the healing length
then phase slippage sets in and the current-phase relation is linear, in contrast to its sinusoidal shape in the Josephson regime.  The transition from phase slips to Josephson regime has been observed in superfluid $^4$He by increasing the healing length with respect to the size of the constriction \cite{Hoskinson2006}.

In BECs, Josepshon junctions are experimentally achieved by confining the atoms in an external double-well potential. In these systems the observation of plasma and self-trapping oscillations has been reported  \cite{Albiez2005}, as well as the ac and dc Josephson effects \cite{Levy2007}. 
Also, Josephson oscillations in an array of junctions have been addressed \cite{Cataliotti2001}, demonstrating phase coherence between them.
On the other hand, phase-slips in BECs have been recently investigated both theoretically \cite{Piazza2009,Piazza2010} and experimentally \cite{Ramanathan2010} as a dissipation mechanism of the superfluid flow in toroidally confined condensates with a constriction produced by an external laser beam. 

Constrictions to the superfluid flow need not be created by an external potential, but can also be brought about by the condensate itself.
In particular, in a dipolar Bose-Einsten condensate (dBEC) confined in a toroidal trap, the anisotropic character of the dipolar interaction induces two potential barriers along the torus, which can be regarded as constrictions \cite{Abad2010}. They can be achieved by polarizing the dipoles perpendicularly to the trap symmetry axis and tuning the scattering length to small values.
Under these conditions, a self-induced double-well effective potential appears in the system, due to the combination of the toroidal trapping potential and the mean-field dipolar interaction. 
In the self-induced double-well, a dBEC 
behaves as a self-induced Josephson junction (SIJJ) \cite{Abad2010b}, showing both Josephson and self-trapping (ST) regimes. 
In this letter we study in detail the self-trapping regime
and show that it is closely related to phase-slip dynamics.

In the mean-field framework, the time evolution of the system is given by the 3D time-dependent Gross-Pitaevskii (TDGP) equation:
\begin{eqnarray}
i\hbar\frac{\partial\Psi}{\partial t} =
 \left[ -\frac{ \hbar^2}{2m} \nabla^2 + V_{\text{t}}(\mathbf{r})
+ g|\Psi|^2 +  V_{\text{d}}(\mathbf{r},t) \right] \Psi\,,
\label{tdgp}
\end{eqnarray}
where $\Psi\equiv\Psi(\mathbf{r},t)$ is the condensate wave function, $m$ is the atomic mass, and $g=4\pi\hbar^2 a/m$ is the coupling constant of the contact interaction, with $a$ the {\it s}-wave scattering length.
The toroidal confinement is simulated by a harmonic plus Gaussian potential:
$ V_{\text{t}}(\mathbf{r})= (m/2)\,\omega_\perp^2(r_\perp^2 + \lambda^2 z^2)
   + V_0 \,
\exp ( -2 \, r_\perp^2/\;\sigma_0^2)\,,
$
where $\omega_\perp$ is the radial harmonic trap frequency, $\lambda=\omega_z/\omega_\perp$ is the trap aspect ratio, with $\omega_z$ the
frequency in the $z$ direction, and $V_0$ and $\sigma_0$ are the intensity and waist of the laser beam producing the central hole of the torus.
For a condensate of polarized magnetic dipoles, the mean-field dipolar potential can be written as
\begin{equation}
  V_{\text{d}}(\mathbf{r},t)=
\frac{\mu_0 \mu^2}{4 \pi}\int d\mathbf{r^\prime}
\frac{1 - 3 \cos^2 \theta}{|\mathbf{r}-\mathbf{r'}|^3} |\Psi(\mathbf{r^\prime},t)|^2 \,,
 \label{dip-pot}
\end{equation}
where $\mu_0$ is the vacuum permeability,
$\mu$ the magnetic moment of the atoms,
and $\theta$ is the angle between  $\textbf{r} - \textbf{r}^\prime$
and the magnetization axis.
The wave function $\Psi$ is found as a function of time by evolving Eq.(\ref{tdgp}) in real time. 
We apply Hamming's algorithm (predictor, corrector, modifier) initialized with a fourth-order Runge-Kutta method, and treat the dipolar term using Fourier transform techniques (see \cite{Abad2010b} for further details).

In the calculations we consider a condensate with $N=5\times10^4$ atoms of $^{52}$Cr
with $a=14\, a_B$ and $\mu=6\mu_B$ ($\mu_B$ is the Bohr magneton),
confined in a toroidal potential with
$\omega_\perp=8.4\times2\pi$ s$^{-1}$, $\lambda=11$, $V_0=30\,\hbar\omega_\perp$ and
$\sigma_0=2.08 \,a_\perp$, where $a_\perp=\sqrt{\hbar/m\omega_\perp}$ 
is the radial harmonic oscillator length.  Since the system is scalable, it can be mapped into other trap parameters and bosonic species \cite{Abad2010b}.  
We want to stress that the trap symmetry axis lies along the $z$ axis, whereas the dipoles are oriented perpendicularly to it, 
along the $y$ axis.
In this configuration the effective potential $V_{\text{eff}}(\mathbf{r},t)=V_{\text{t}}(\mathbf{r})+V_{\text{d}}(\mathbf{r},t)$ is a ring-shaped double well with the two wells on the $x$ axis. 
Note that this effective potential depends on time through the condensate wave function in the dipolar term. 
The self-induced double-well structure is not exclusive of dipolar condensates \cite{Abad2010}, but has also been predicted to appear in degenerate dipolar Fermi gases \cite{Dutta2006}, and recently in more general dipolar Bose and Fermi  systems \cite{Zollner2010}.
However, coherent tunneling dynamics is only possible if the system is in the superfluid regime.

Dynamics in the SIJJ consists on the coherent tunneling of atoms between the two weakly linked dBECs. It is generally well captured by two conjugate variables: the population imbalance, $ Z(t)= (N_L(t)-N_R(t)) / N$, and the phase difference, $ \phi(t)=\phi_R(t)-\phi_L(t)\label{phase}$, where $N_{L(R)}(t)$ corresponds to the number of atoms on the left (right) well,
and $\phi_{L(R)}(t)$ to the phase of the dipolar condensate averaged on the left (right) well. Note that this description assumes, in particular, that the phase of the wave function is uniform in each well. For narrow junctions this is a good approximation, but it starts failing for wider junctions.

For initial conditions $Z(0)=0.67$ and $\phi(0)=0$ the system is in the self-trapping regime. 
This regime is  brought about by the nonlinear interactions in the condensate  \cite{Smerzi1997} and is characterized
by  a nonzero time average of the population imbalance, that is, the trapping
of the most part of the atoms in one of the wells. It is reached for strong enough interactions and for a large initial population imbalance between the two wells.
In the cases observed experimentally \cite{Albiez2005}, ST was accompanied by a linear increase of the phase difference between the two weakly linked condensates, characterizing thus a running-phase mode.

\begin{figure}[h]
\epsfig{file=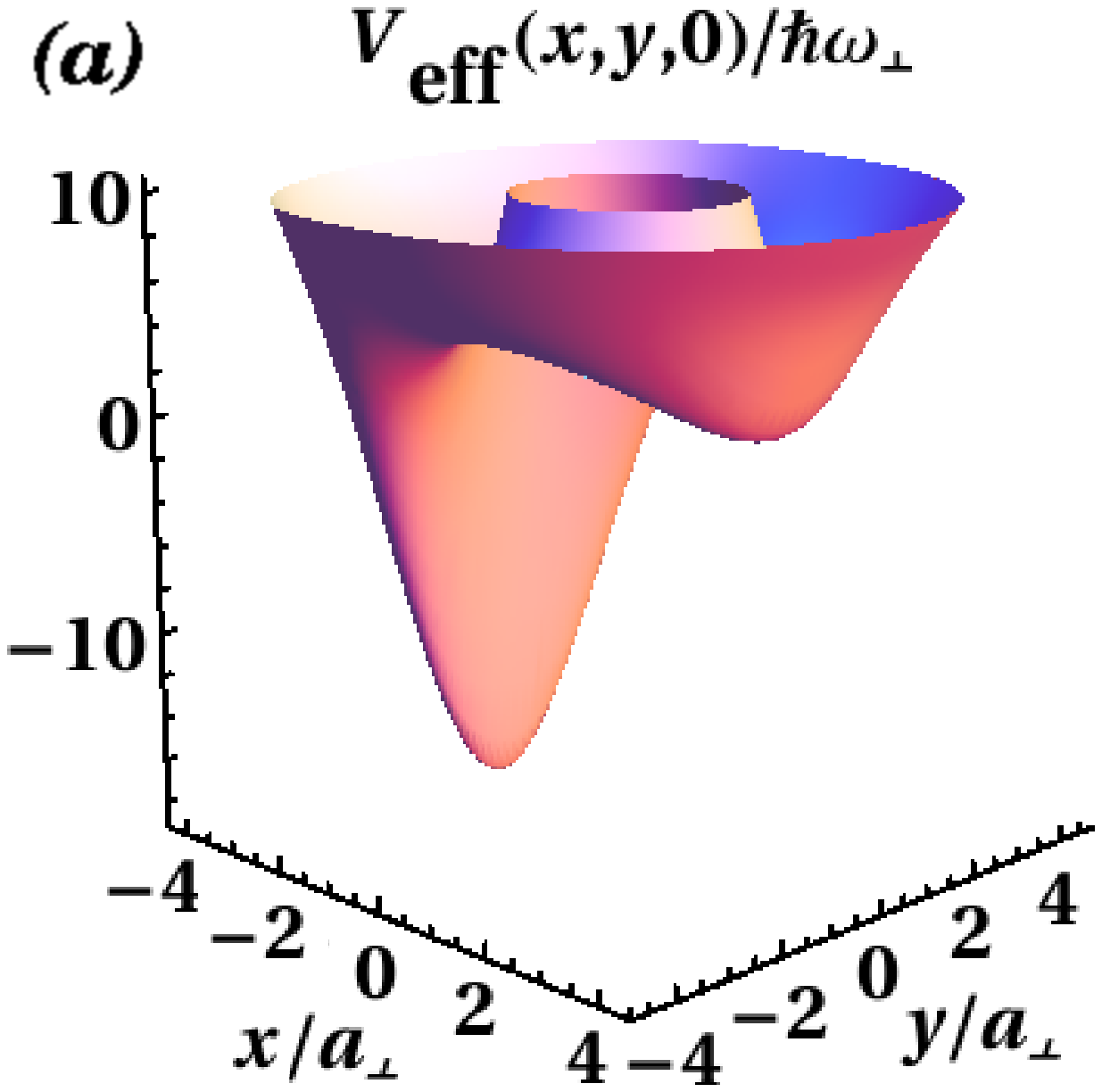, width=0.45\linewidth, clip=true}
\epsfig{file=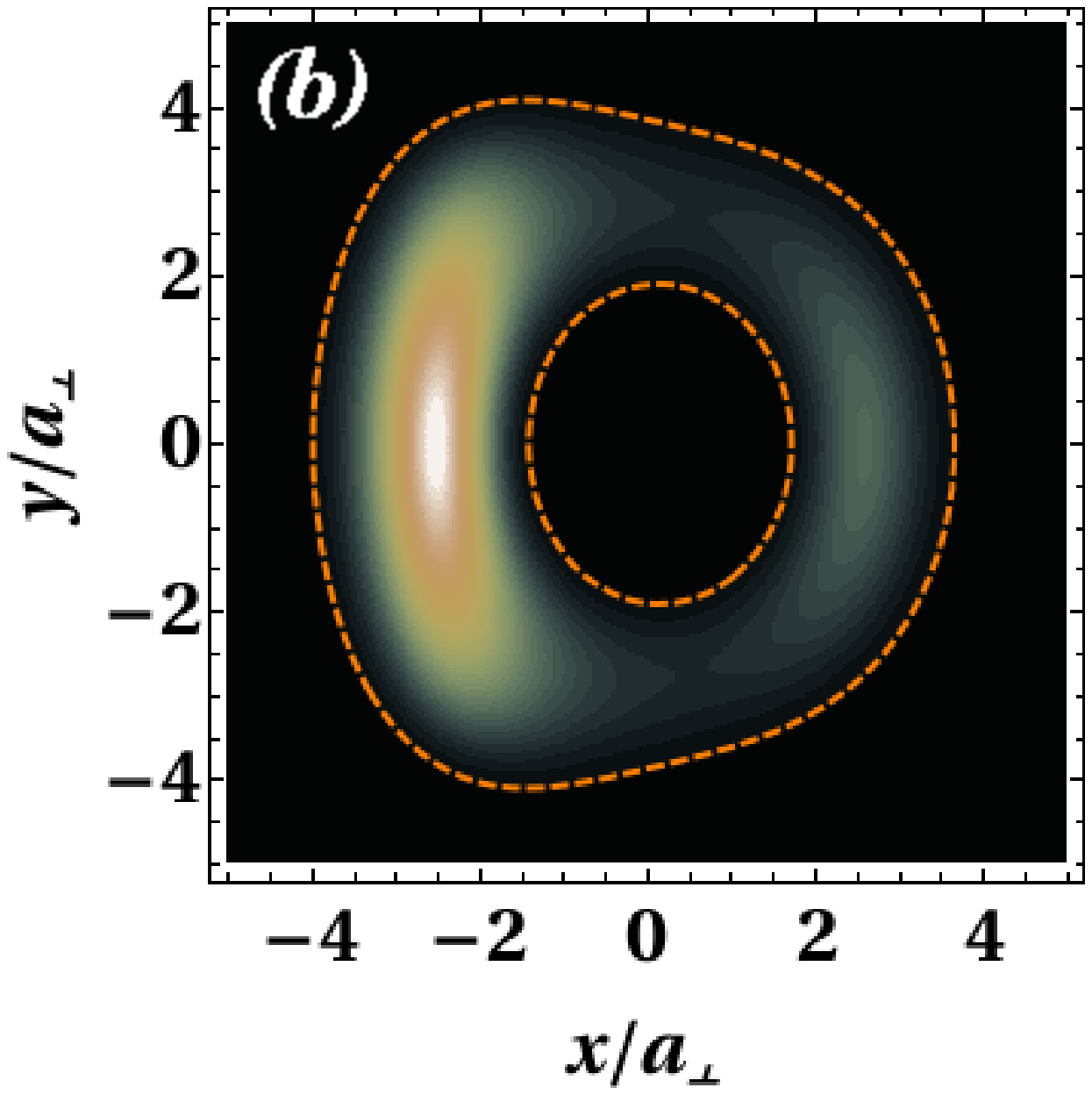, width=0.45\linewidth, clip=true}
\epsfig{file=fig1c.eps, width=0.9\linewidth, clip=true}
\caption{
(Color online) Initial effective potential (a) and corresponding density (b) in the plane $z=0$.
Imbalance and phase difference as a function of time.
}
\label{fig1}
\end{figure}

Figure~\ref{fig1}(a) shows the effective potential at $t=0$, in the $z=0$ plane. It has the shape of a double well in the azimuthal direction, with two potential barriers at $y>0$ and $y<0$, which we will name as upper and lower junction, respectively.
The two junctions are coupled in phase, so they can be effectively regarded as only one,
analogously to
the array of nanoapertures in helium experiments (see, for instance, \cite{Hoskinson2006}). The initial density structure with the two links is shown in Fig.~\ref{fig1}(b).
As a guide to the eye the dashed line draws the equidensity contour at $0.1 \rho_{\rm max}$, being $\rho_{\rm max}$ the maximum density in the plane $z=0$.
We have taken  into account in the definition of left and right wells the displacement
of the junctions around $x\simeq a_\perp$ with an inclination of about $15^\circ$ from the $y$ axis. This gives rise to a left well slightly larger than the right one \cite{footnote1005}.

Fig.~\ref{fig1}(c) shows the coupled dynamical evolution of $Z(t)$ and $\phi(t)$
for the above initial conditions.
The time derivative of the population imbalance is related to the flow of atoms across the junction.
At the maxima and minima of the imbalance its slope changes sign, so that the flow of atoms is reversed.
In a two mode description, when the imbalance reaches a maximum the phase difference is zero, while in a minimum it reaches $\pi$ \cite{footnote1}.
%
However, looking more closely to the coupled dynamics given by TDGP, we see that in the minima of imbalance the phase difference does not exactly correspond to $\phi=\pi$. This fact hints at a richer dynamics that is not accounted for in the two macroscopic variables $Z$ and $\phi$, but comes from a local behavior. 
In other words, since the junction is a wide junction, the local currents within it affect the gross dynamics contained in $Z$ and $\phi$.
To understand better the irregularities in the signal of Fig.~\ref{fig1}(c), we have to look at the dynamics of the local phase.
We find two main results. 
On the one hand, we see that the flux inversion in the minima of the imbalance is accompanied by the crossing of a vortex through each junction. These vortices are dynamically created and constitute the phase-slip process that takes place when the phase difference reaches $\pi$.
On the other hand, the currents generated by the phase gradients within the junction propagate and affect the phase coherence in each one of the wells, giving rise to a spatial dependence of the phase. This new phase gradients become currents in the wells that are not properly accounted for in $Z$ and $\phi$ and are mostly the  responsible of the dephasing between these two variables.

\begin{figure}[h]
\epsfig{file=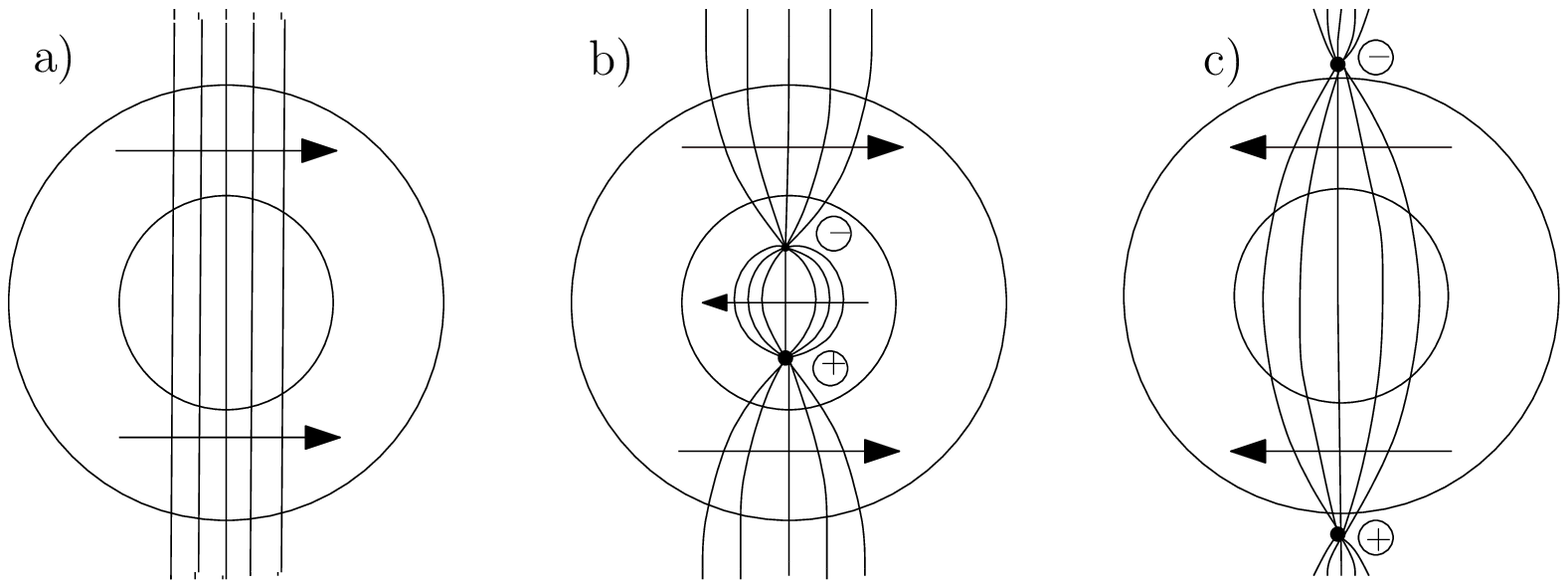, width=\linewidth, clip=true}\vspace{1em}
\epsfig{file=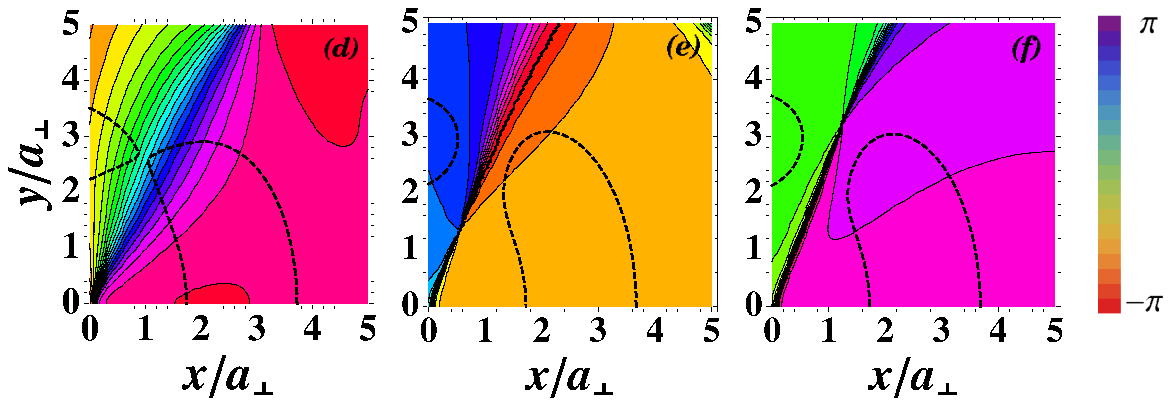, width=\linewidth, clip=true}
\caption{ (Color online) A scheme of the  slippage process is given in the top panel.
The solid lines represent constant phase curves, the dashed lines a torus contour in the $z=0$ plane.
The $ + $ ($-$) signs the position of a vortex (antivortex),
and  the arrows indicate
 the phase gradient, and thus the velocity field,  along the junctions.
We have drawn the following three steps: a) Phase accumulation along the junctions,
b) Vortex-antivortex pair
separation after the nucleation and c) Inversion of the velocity field after the vortices cross  the torus.
In the bottom panel we show snapshots of the numerical phase
around the upper junction at $z=0$,
 as a function of $x$ and $y$.
The times are: (d) $ t= 6.41\, \omega_\perp^{-1} $, (e) $ t= 8.12\, \omega_\perp^{-1} $,  and (f)
 $ t= 8.76\, \omega_\perp^{-1} $.
}
\label{fasest}
\end{figure}

Let us  examine the dynamical nucleation of vortices in more detail.
In the top panels of Fig.~\ref{fasest} we schematically describe the vortex-induced phase-slip process
that contributes
to the flux inversion in the minima of the imbalance.
Before reaching the minimum, since the phase difference is very large, there is an accumulation of
phase between the two condensates, producing a large gradient of the phase along
the line that connects both junctions.
The atoms tunnel from the left to the right well, acquiring a velocity associated to this gradient, as indicated in Fig.~\ref{fasest}(a).
The velocity is larger in the low density regions, namely the central hole giving rise to the toroidal geometry and the external region.
When the phase difference is $\pi$, the velocity reaches a critical value and two vortices are nucleated,
a vortex-antivortex pair
to ensure angular momentum conservation, as seen in Fig.~\ref{fasest}(b) \cite{footnote2}.
Subsequently both vortices cross simultaneously the section of the torus in opposite directions,
each one moving outwards through one link, producing the phase slip in the junction.
After the passage of the vortex (antivortex) the sign of the local velocity in the link is changed, leading
thus to the inversion of the flux of atoms, as shown
schematically in Fig.~\ref{fasest}(c).
The vortex (antivortex) with positive (negative) charge is defined with the phase gradient
in the counterclockwise (clockwise) direction.

Panels (d), (e) and (f) of Fig.~\ref{fasest} show
the phase of the wave function, in the $z=0$ plane of  the upper junction,
around the first minimum of $Z(t)$ at the moments schematically represented in (a), (b) and (c), respectively.
A vortex-antivortex pair is nucleated in the center, separates and then the vortex (antivortex) crosses the lower
(upper) junction.
The same effect is produced if an antivortex and a vortex are nucleated at opposite points of the external surface of the condensate and then cross the torus and anihilate at the center.
Analyzing the dynamics we see that both processes take place at different minima.
A similar phenomenon has been addressed in Ref.~\cite{Piazza2009}, where a single external barrier was raised in a toroidally confined $s$-wave condensate
with a nonzero initial angular momentum. In that work the appearance of vortices 
from the external or internal surfaces of the torus was controlled by the height of the barrier, while the nonzero angular momentum favored  the first process. In the case reported in this article, however, both processes are equivalent.
We want to note that all the vortices we predict correspond to quantized vortex lines, with no visible bending along the $z$ direction. The quantization is guaranteed by the fact that the phase changes by $2\pi$ around the singularity.
The angular momentum remains zero throughout the simulation and there is no net creation of vorticity in the system.

\begin{figure}[h]
\epsfig{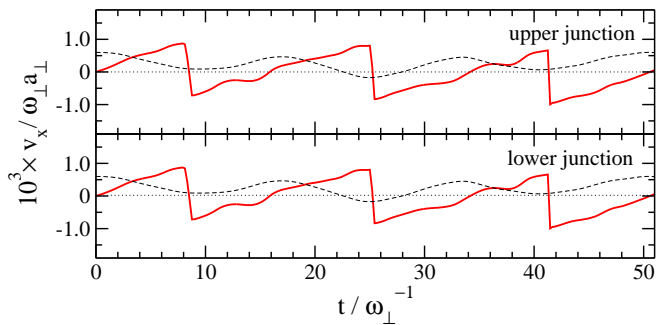}
\caption{(Color online) Velocity of the atoms at the upper and lower junctions as a function of time. The steep drops correspond to the crossing of a vortex (phase-slip). As a guide to the eye, the imbalance has been plotted in an arbitrary scale as a dashed line (see Fig.\ref{fig1}).
}
\label{velo}
\end{figure}

In order to complete the description of the vortex-induced phase slip, we have computed the mean velocity of the atoms across the upper and lower junction as a function of time, see Fig.~\ref{velo}.
As expected, both junctions behave in phase and present the same physics:
simultaneously to the nucleation and crossing of
the vortices through the junctions, there is a steep drop of the velocity in
each junction. In analogy with the experiments carried out in superfluid helium, this drop marks the point at which a phase slip takes place.
From the sharp decrease of the mean velocity it is possible to estimate the vortex passage velocity using
$v_p=\Delta/\delta t$, where $\Delta$ is the length of the barrier and $\delta t$ is the time it takes the
vortex to cross it as read from Fig.~\ref{velo}. For the three minima it gives: $v_p^{(1)}=2.7\,\omega_\perp a_\perp$, $v_p^{(2)}=4.2\,\omega_\perp a_\perp$ and $v_p^{(3)}=8.2\,\omega_\perp a_\perp$.

In conclusion we have shown that there is a close relationship between
phase-slips and self-trapping dynamics in a ring-shaped Josephson junction.
The coherent passage of a vortex and an antivortex across the two coupled junctions triggers the inversion of the flux of atoms in the minima of imbalance, where the phase difference reaches $\pi$.
These results are the first step to bring together the processes of self-trapping, ac Josephson effect and phase-slippage and to gain a deeper understanding of non-linear dynamics in bosonic Josephson junctions.

We believe that these physics can be experimentally addressed, since the needed techniques
are nowadays available: toroidal traps are
currently constructed in actual experiments \cite{ryu07,torus}.
The initial imbalance can be achieved
using either an external ramp or by shifting the laser beam. The reduction of the
scattering length is provided by the utilization of Feshbach resonances procedures.
Our results involve Cr atoms \cite{expCr}, but also Li or K are candidates to be experimentally
studied \cite{alkali}. Both density and phases have been measured in experiments \cite{Albiez2005}.
Experimental observation of vortex-dipoles in Bose-Einstein condensates \cite{Neely2010} has also been
addressed.

Finally we want to point out that, although we have studied the case of a SIJJ, the relation between self-trapping and vortex dynamics is not restricted to dipolar condensates but it
can be extended to any other type of bosonic Josephson junctions \cite{wip}.

The authors would like to thank S. Shenoy for helpful discussions.
We acknowledge financial support under Grant No. FIS2008-00421 from MEC (Spain), Grant No. 2009SGR1289 from Generalitat de Catalunya (Spain), and Grant No.
PIP 11420090100243 from CONICET (Argentina).
M. A. is supported by CUR (Generalitat de Catalunya).

\end{document}